\documentstyle[preprint,floats,tighten,aps,graphicx,epsf]{revtex}

\newif\ifpdf
\ifx\pdfoutput\undefined
\pdffalse % we are not running PDFLaTeX
\else
\pdfoutput=1 % we are running PDFLaTeX
\pdftrue
\fi

\def\bsigma{\mbox{\boldmath $\sigma$}}

\preprint{\vbox{ \hbox{CMU-HEP01-04}   \hbox{FERMILAB-Pub-01/098-T} 
\hbox{UCSD/PTH 01-08} }}
\title{\vspace{.5cm}Resumming the color-octet contribution to radiative 
$\Upsilon$ decay} 
\author{Christian W. Bauer $^a$, Cheng-Wei Chiang $^b$, Sean Fleming $^b$,\\
\vspace{.1cm} Adam K. Leibovich $^c$, and Ian Low $^b$}
\address{ \vbox{\vskip 0.5truecm} 
 $^a$ Department of Physics, University of California at San Diego,
       La Jolla, CA 92093 \\ 
 \vspace{.3cm}
 $^b$ Department of Physics, Carnegie Mellon University,
      Pittsburgh, PA 15213 \\
 \vspace{.3cm}
 $^c$ Theory Group, Fermilab, P.O. Box 500, Batavia, IL 60510}

\begin{document}

%%%%%%%%%%%%%%%%%%%%%%%%%%%%%%%%%%%%%%%%%%
%Some more stuff to get graphics to work
\ifpdf
\DeclareGraphicsExtensions{.pdf, .jpg}
\else
\DeclareGraphicsExtensions{.eps, .jpg,.ps}
\fi
%%%%%%%%%%%%%%%%%%%%%%%%%%%%%%%%%%%%%%%%%%

%%%%%%%%%%%%%%%%%%%%%%%%%%%%%%%%%%%%%%%%%%
%Create the title page
\maketitle
\begin{abstract}

At the upper endpoint of the photon energy spectrum in $\Upsilon \to X
\gamma$, the standard NRQCD power counting breaks down and the OPE
gives rise to color-octet structure functions. Furthermore, in this
kinematic regime large Sudakov logarithms appear in the octet Wilson
coefficients.  The endpoint spectrum can be treated consistently
within the framework of a recently developed effective field theory of
collinear and soft particles. Here we show that within this approach
the octet structure functions arise naturally and that Sudakov
logarithms can be summed using the renormalization group equations.
We derive an expression for the resummed energy spectrum and, using a
model lightcone structure function, investigate the phenomenological
importance of the resummation.

\end{abstract}

%%%%%%%%%%%%%%%%%%%%%%%%%%%%%%%%%%%%%%%%%%
\tighten
\newpage
%%%%%%%%%%%%%%%%%%%%%%%%%%%%%%%%%%%%%%%%%%
%Main body of the paper

\section{Introduction}

Early theoretical analyses of heavy quarkonium decay were based on
the color-singlet model (CSM). The underlying assumption of this model
is that the heavy-quark--antiquark pair has the same quantum numbers as
the quarkonium meson. (For example the $b\bar{b}$ that forms an
$\Upsilon$ must be in a color-singlet ${}^3S_1$ configuration.) One
consequence of such a restrictive assumption is that theoretical
predictions based on the CSM are simple, depending on only one
nonperturbative parameter. The quantities first calculated in the CSM
were the inclusive rates for quarkonium to decay into leptons and
into light hadrons~\cite{first}. Subsequently the direct photon
spectrum in inclusive radiative quarkonium
decays was calculated~\cite{firstRad}.

In recent years the simple CSM has been superseded by a
nonrelativistic effective theory of QCD
(NRQCD)~\cite{bbl,lmr}. Inclusive decays of quarkonium are now
understood in the framework of the operator product expansion (OPE),
supplemented by the power-counting rules of NRQCD. In this formalism
the direct photon spectrum of $\Upsilon$ decay is 
\begin{equation}
\label{ope}
\frac{d \Gamma}{d z} = \sum_i C_i(M,z) 
\langle \Upsilon | {\cal O}_i | \Upsilon \rangle \,, 
\end{equation}
where $z = 2 E_\gamma / M$, with $M = 2 m_b$. The $C_i$ are
short-distance Wilson coefficients which can be calculated as a
perturbative series in $\alpha_s(M)$, and the ${\cal O}_i$ are NRQCD
operators. NRQCD power-counting rules assign a power of the relative
velocity $v$ of the heavy quarks to each operator and organizes the
series. The series may be truncated at any order with omitted terms 
suppressed by powers of $v$. For $S$-wave mesons the formally leading-order
contribution is the color-singlet operator, which is related to the
wavefunction at the origin. Thus for $S$-wave decays one recovers the
CSM at leading order in $v$. At higher orders in $v$ color-octet
operators need to be included. 

%Though they are subleading, they can be
%important in some processes because the color-octet short distance
%coefficients may be lower order in $\alpha_s(M)$.
%The photon spectrum in $\Upsilon \to X \gamma$ decay has been studied
%in the framework of NRQCD and the 
%OPE~\cite{Maltoni:1999nh,Wolf:2001pm,upsStructfun}. 

However, the picture of the photon spectrum in $\Upsilon \to X \gamma$
decay which emerges is much richer than the naive expectation that the
color-singlet contribution is
leading~\cite{Maltoni:1999nh,Wolf:2001pm,upsStructfun}. At low values
of the photon energy, fragmentation contributions to $\Gamma(\Upsilon
\to X \gamma)$ are important \cite{phofrag,Maltoni:1999nh}. The
situation at large values of the photon energy is even more
interesting, because both the OPE and perturbative expansion break
down. The breakdown of the OPE was first addressed in
Ref.~\cite{upsStructfun}. It was shown that the color-octet
contributions, which give rise to a singular contribution at maximum
photon energy, become leading for large photon energies. The singular
nature is smeared by a nonperturbative structure function, which tames
the endpoint behavior of the photon spectrum.  The breakdown of the
perturbative expansion gives rise to so-called Sudakov logarithms
which have to be resummed. In a recent work~\cite{Hautmann:2001yz} it
was pointed out that the leading Sudakov logarithms cancel in the CSM.
However this is not the case for Sudakov logarithms in the color-octet
contribution.

Both the breakdown of the OPE and the appearance of Sudakov
logarithms are symptoms of the same disease: NRQCD does not contain
the correct low energy degrees of freedom to describe the endpoint of
the photon spectrum. It does not contain collinear quarks and
gluons. A theory constructed from the appropriate degrees of
freedom was developed in Refs.~\cite{csetI,csetII}. In those papers the
theory was applied to the decay of a single heavy quark to light
degrees of freedom. It was shown that the renormalization group
equations (RGEs) in this theory sums Sudakov logarithms. In addition,
for inclusive decays at the endpoint, the nonperturbative structure
function arises naturally from a modified version of the OPE. Here we
apply the theory to the color-octet contributions to radiative
$\Upsilon$ decay. In Section II we discuss the leading contributions
in the endpoint region and motivate perturbative and nonperturbative
resummation. In Section III we sum Sudakov logarithms using the RGEs
in an effective field theory. In Section IV we introduce a
phenomenological model for the shape function, convolute it with the
resummed spectrum, and show how this changes the color-octet
contribution to the spectrum. In Section V we conclude.  

\section{Leading order results}

The inclusive radiative differential decay rate of $\Upsilon$ can be
calculated using the optical theorem.  This relates the decay rate to
the imaginary part of the forward matrix element of the time ordered
product of two currents
\begin{equation}
\label{diffrate}
\frac{d \Gamma}{d z} = \frac{M^2}{8 \pi^2} \, z \,
\langle \Upsilon | {\rm Im}\, T | \Upsilon \rangle \,,
\end{equation}
where we have used
nonrelativistic normalization for the states: $ \langle \Upsilon (P')
| \Upsilon (P) \rangle = (2 \pi)^3 \delta^3(P'-P)$.  For large momentum transfer the time ordered
product can be expanded in terms of local operators giving
\begin{eqnarray}
\label{ope_rate}
\frac{d \Gamma}{d z} = \sum_i C_i(z) \langle \Upsilon | 
{\cal O}_i | \Upsilon  \rangle\,.
\end{eqnarray}
The Wilson coefficients, $C_i(z)$, can be calculated perturbatively as
a series in $\alpha_s(M)$. The parametric size of the long distance
matrix elements are determined by power counting in NRQCD, but to
obtain quantitative results these matrix elements have to be extracted
from experiments or lattice calculations.

At leading order in the NRQCD $v$ expansion, only the color-singlet,
spin-triplet operator
\begin{eqnarray}
\label{csme}
{\cal O}_{\bf 1}(^3S_1) = \sum_{{\bf p}, {\bf p}'}
[\psi^\dagger_{{\bf p}'} \sigma^i \chi_{-{\bf p}'}] \;
[\chi^\dagger_{-{\bf p}} \sigma^i \psi_{{\bf p}} ]
\end{eqnarray}
contributes, with the leading order Wilson coefficient
\begin{eqnarray}
\label{cswc}
C_{\bf 1}^{(0)}(^3S_1)(z) &=& 
\frac{128\,\alpha_s^2\,\alpha\,e_Q^2}{27 M^2}\,\Theta(1-z)
\nonumber \\
&& \times
\left[ \frac{2-z}{z} + \frac{z(1-z)}{(2-z)^2} + 2 \frac{1-z}{z^2} 
\log(1-z) - 2 \frac{(1-z)^2}{(2-z)^3} \log(1-z) \right]\,,
\end{eqnarray}
where $e_Q=-1/3$ for $\Upsilon$.
This is the CSM result~\cite{firstRad}, with $\langle \Upsilon |
{\cal O}_{\bf 1}({}^3S_1)|\Upsilon \rangle \approx (3/2 \pi) |R(0)|^2$,
where $R(0)$ is the radial wavefunction at the origin. 
The first color-octet contributions to (\ref{ope_rate}) are suppressed
by $v^4$ relative to this color-singlet contribution.  There are two
operators
\begin{eqnarray}
\label{co_ops}
{\cal O}_{\bf 8}(^1S_0) &=& \sum_{{\bf p},{\bf p}'}
[\psi^\dagger_{{\bf p}'}\, T^a\, \chi_{-{\bf p}'}] \; 
[\chi^\dagger_{-{\bf p}}\, T^a\, \psi_{{\bf p}}] \,,
\nonumber\\
{\cal O}_{\bf 8}(^3P_0) &=& \frac{1}{3} \sum_{{\bf p},{\bf p}'}
[\psi^\dagger_{{\bf p}'}\, {\bf p}' \cdot \bsigma\, T^a \chi_{-{\bf p}'}] \;
[\chi^\dagger_{-{\bf p}}\, {\bf p} \cdot \bsigma\, T^a \psi_{{\bf p}}] \,,
\end{eqnarray}
with leading order Wilson coefficients\cite{Maltoni:1999nh} 

\begin{eqnarray}
C_i^{(0)}(z)=\tilde{C}_i^{(0)} \delta(1-z)\,, \label{samecoeff}
\end{eqnarray}
 where
\begin{eqnarray}
\label{cowc}
\tilde{C}_{\bf 8}^{(0)}(^1S_0) &=& 
  \frac{16\, \alpha_s\,\alpha\, e_Q^2 \pi}{M^2}, \nonumber \\
\tilde{C}_{\bf 8}^{(0)}(^3P_0) &=& 
  \frac{448\, \alpha_s\,\alpha\, e_Q^2 \pi}{M^4} \,.
\end{eqnarray}
Since the color-octet Wilson coefficients are enhanced by a power of
$\alpha_s(M)/\pi$ relative to the color-singlet one, the overall
suppression of the color-octet contribution is $\pi v^4/\alpha_s(M)
\sim v^2$, where we have used that numerically $\alpha_s/\pi \sim
v^2$.

The singular nature of the coefficients (\ref{samecoeff}) is an
indication that the OPE is breaking down.  We can obtain a rough
estimate for the value of $z$ at which the octet contributions become
of order the color-singlet contribution by smearing the perturbative
spectrum over a small region near $z=1$. Integrating over $1-v^2 < z <
1$ gives a color-singlet contribution that scales as $\alpha^2_s(M)
v^2$ and a color-octet contribution that scales as $\alpha_s(M) \pi
v^4$. Thus, the ratio of octet to singlet in this region of phase
space is $\pi v^2 /\alpha_s(M) \sim 1$, making the color-octet
contribution of the same order as the color-singlet one.

It was shown in Ref.~\cite{upsStructfun} that in precisely this
endpoint region the OPE breaks down and an infinite set of operators
have to be resummed into lightcone distribution functions
$f_i(k_+)$.\footnote{This is very similar to the behavior of the OPE
for $B \to X_u \ell \bar{\nu}$ at the endpoint of the lepton energy
spectrum and $B \to X_s \gamma$ at the endpoint of the photon energy
spectrum~\cite{Bstructfun}.} Each structure function gives the
probability to find a $b \bar b$ pair with the appropriate quantum
numbers and residual momentum $k_+$ in the $\Upsilon$. For the
color-singlet contribution the structure function can be calculated
using the vacuum saturation approximation. It simply shifts the
maximal photon energy from $2 m_b$ to $M_\Upsilon$
\cite{upsStructfun}. The color-octet contributions, however, give rise
to two new nonperturbative functions at leading order. At higher order
there are an infinite number of additional structure functions, so the
differential decay rate in the endpoint region is
\begin{eqnarray}
\label{shapefunc}
\frac{d \Gamma}{d z} = \sum_i \int dk_+ C_{i}(z,k_+) 
f_{i}(k_+) \langle  \Upsilon | {\cal O}_{i} | \Upsilon \rangle\,.
\end{eqnarray}

Another effect one encounters in the endpoint region is the appearance
of Sudakov logarithms in the Wilson
coefficients, which ruin the perturbative expansion. Consider, for
example, the Wilson coefficients for the
color-octet operators at next-to-leading order in $\alpha_s$. In the $z\to 1$ limit they are~\cite{Maltoni:1999nh}
\begin{eqnarray}
\label{WCoct1}
C_{\bf 8}^{(1)}(^1S_0)(z) &=& \frac{\alpha_s}{2\pi}\tilde{C}_{\bf 8}^{(0)}(^1S_0)
   \left[-2C_A \left(\frac{\log(1-z)}{1-z}\right)_+ - 
   \left(\frac{23}{6} C_A - \frac{n_f}3\right)\left(\frac1{1-z}\right)_+
   \right] ,
\nonumber \\
C_{\bf 8}^{(1)}(^3P_0)(z) &=& \frac{\alpha_s}{2\pi}\tilde{C}_{\bf 8}^{(0)}(^3P_0)
   \left[-2C_A \left(\frac{\log(1-z)}{1-z}\right)_+ - 
   \left(\frac{23}{6} C_A - \frac{n_f}3\right)\left(\frac1{1-z}\right)_+
   \right] .
\end{eqnarray}
If these coefficients are integrated over the shape-function region
($1-v^2 < z < 1$), then the first plus distribution on the
right-hand side gives rise to a double logarithm, $\log^2 v^2$, while
the second plus distribution gives a single logarithm, $\log v^2$.  Both
of these are
numerically of order $1/\alpha_s$. This clearly ruins the perturbative
expansion. Therefore, to obtain a well controlled expansion, these
logarithms must be summed. 

\section{Summing Sudakov Logarithms}

The NRQCD power-counting rules break down as $z \to 1$ because NRQCD
does not include all of the long distance modes: collinear physics is
missing from the theory. An effective theory which includes collinear
physics was developed in Refs.~\cite{csetI,csetII}. This collinear-soft
theory describes the interactions of highly energetic collinear modes
with soft degrees of freedom. To describe $\Upsilon$ decay at
the endpoint we have to couple the collinear-soft theory with
NRQCD~\cite{lmr}. This is analogous to $B \to X_s \gamma$ decays at
the endpoint, which was studied in the context of effective field
theory in Ref.~\cite{csetI}. We will closely follow the development in
that paper.

Understanding inclusive $\Upsilon \to X \gamma$ decays near the
endpoint is a two-step process.\footnote{In Ref.~\cite{Manohar:2000mx}
it was pointed out that at higher orders in perturbation theory one
has to adopt a one-step scheme similar to the one developed in a
slightly different context in \cite{lmr}.} In the first step we must
integrate out the large scale, $M=2 m_b$, set by the $b\bar{b}$ pair
constituting the $\Upsilon$. This is done by matching onto the
collinear-soft theory. In the second step collinear modes are
integrated out at a scale which is set by the invariant mass of the
collinear jet. This is done by performing an OPE and matching onto a
soft theory containing operators which are nonlocal along the
lightcone and whose matrix elements are the lightcone distribution
functions, $f_i(k_+)$.  Sudakov logarithms are summed by using the
effective theory RGEs. Operators are run from the hard scale $M$ to
the collinear scale where the OPE is performed. The nonlocal
soft operators are then run from the collinear scale down to a soft
scale where their matrix elements do not contain large
logarithms. This procedure sums all Sudakov logarithms into the short
distance coefficient functions.

To better understand the scales involved consider the momentum of a
collinear particle moving near the lightcone. In lightcone coordinates
we can write this momentum as $p=(p^+,p^-,p_\perp)$. Since the mass of
the particle is much smaller than its energy, we define $p^2 \sim M^2
\lambda^2$, where $M$ is the scale that sets the energy and $\lambda$
is a small parameter. The lightcone momentum components are widely
separated. If we choose $p^-$ to be ${\cal O}(M)$, then $p_\perp/p^-
\sim \lambda$, and $p^+/p^- \sim \lambda^2$. We refer to these two
scales as collinear and soft, respectively. To be concrete consider
the $b \bar b$ pair to have momentum $Mv^\mu+k^\mu$, where $v^\mu =
(1,0,0,0)$ and $k^\mu$ is ${\cal O}(\Lambda_{\rm QCD})$ in the
$\Upsilon$ center-of-mass frame. The photon momentum is $Mz\bar{n}^\mu
/2$, where we have chosen $\bar{n}^\mu = (1,0,0,1)$. In the endpoint
region the hadronic jet recoiling against the photon moves in the
opposite lightcone direction $n^\mu = (1,0,0,-1)$, with momentum
$p^\mu_X = Mn^\mu /2 + M(1-z) \bar{n}^\mu /2 + k^\mu$ . Thus
the hadronic jet has $\bar{n}\cdot p_X = p^-_X \sim M$. Next note that
$m^2_X \approx M^2(1-z)$. For $(1-z) \sim v^2 \sim \Lambda_{\rm
QCD}/M$ we find
\begin{eqnarray}
m_X \sim \sqrt{M \Lambda_{\rm QCD}}\,,
\end{eqnarray}
which is the collinear scale. This implies that for this process
the collinear-soft expansion parameter $\lambda$ is of order
$\sqrt{1-z}\sim\sqrt{\Lambda_{\rm QCD}/M}$. 
The soft scale is the component of the
hadronic momentum in the $n$ direction:
\begin{eqnarray}
n \cdot p_X \sim \frac{m^2_X}{\bar{n} \cdot p_X} 
\sim \Lambda_{\rm QCD} \sim M \lambda^2\,.
\end{eqnarray}
Thus in order to sum Sudakov logarithms in $\Upsilon \to X \gamma$ we
first match onto the collinear-soft theory at $M$ and run operators
in this theory down to the collinear scale $\mu_c \sim M\sqrt{1-z}$. At that
scale we perform the OPE by matching onto a soft theory containing
operators that are nonlocal along the lightcone and run these
operators down to the soft scale $\mu_s \sim M(1-z)$.

\subsection{The collinear-soft theory}

\begin{figure}[t]
\centerline{\includegraphics[width=6.5in]{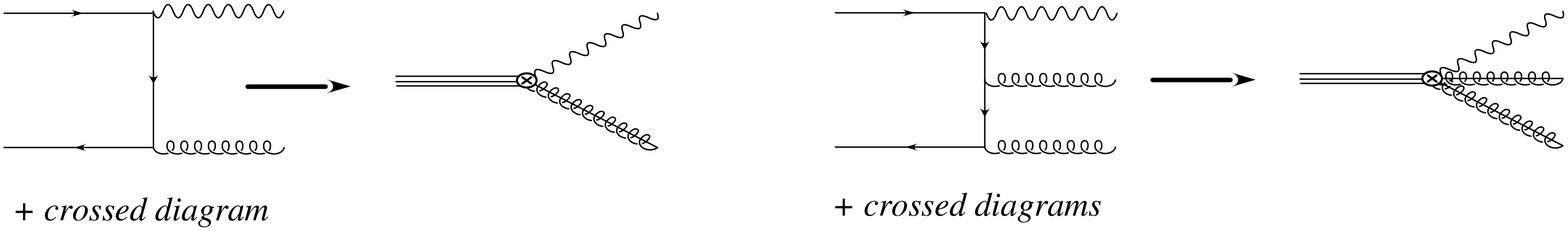}}
\vspace{2em}
\caption{\it Matching onto operators in the effective field theory with one and two gluons in the final state.}
\label{match_leading}
\end{figure}

We first need to integrate out the large scale $M$ by matching onto
the collinear-soft theory.  This is done by calculating matrix
elements in QCD, expanding them in powers of $\lambda$, and matching
onto products of Wilson coefficients and operators in the effective
theory. For the process of interest this matching is illustrated in
Fig.~\ref{match_leading}.
The heavy-quark spinors and the heavy-quark propagator in QCD can be
expanded in powers of $v$ to match onto NRQCD. The QCD
spinor is decomposed into two two-component Pauli spinors 
$\psi$ and $\chi$ for the
heavy quark and anti-quark, respectively \cite{bc}.\footnote{Note that
in Ref.~\cite{bc} states are normalized relativistically.} We also need to
expand the amplitude in powers of $\lambda$ to match onto the collinear-soft
theory. This is done by using the power-counting rules for the gluon
field given in Ref.~\cite{csetII} and by scaling the different
components of the gluon momentum as
\begin{eqnarray}
p_g^\mu &=& p_g\cdot\bar n \,\frac{n^\mu}{2} + p_{g\perp}^\mu + 
  p_g\cdot n \,\frac{\bar n^\mu}{2} \nonumber\\
&=& {\cal O}(1) + {\cal O}(\lambda) + {\cal O}(\lambda^2) \,.
\end{eqnarray}
%To combine the NRQCD expansion with the lightcone expansion, we use
%\begin{eqnarray}
%v \sim \lambda\,.
%\end{eqnarray}
Omitting the straightforward but unenlightening details, the color-octet
contributions match in the $\Upsilon$ rest frame, at leading order in
$\lambda$, onto the operators
\begin{eqnarray}
\label{eff_vertx}
Q_{\bf 8}^\mu(^1S_0) &=& 2ig_s e_b \, 
\epsilon^{\alpha\mu}_\perp \; \sum_{\bf p}
[\chi^\dagger_{-{\bf p}} T^a  \psi_{\bf p}]\, A^a_\alpha\,,
\nonumber \\
Q_{\bf 8}^\mu(^3P_J) &=& 4g_s e_b \left(g_\perp^{\alpha\delta} g^{\mu\sigma} +
g_\perp^{\alpha\sigma} g^{\mu\delta} -
g_\perp^{\alpha\mu} \bar{n}^\sigma \bar{n}^\delta\right)
\; \sum_{\bf p} \Lambda\cdot\widehat{\bf p}_\sigma  
[\chi^\dagger_{-{\bf p}} \Lambda\cdot\sigma_\delta
T^a  \psi_{\bf p}]\,A^a_\alpha \,,
\end{eqnarray}
where $\epsilon^{\alpha\mu}_\perp =
\epsilon^{\alpha\mu\rho\beta}\bar{n}_\rho v_\beta$ and
$g_\perp^{\mu\nu} = g^{\mu\nu} - (n^\mu\bar n^\nu + n^\nu\bar
n^\mu)/2$. Hatted variables are divided by $M$, and the $\Lambda^{\mu
i}$ are boost matrices which boost from the $b\bar{b}$ center-of-mass
frame~\cite{bc}. We have chosen factors in the operators such that at
leading order the corresponding Wilson coefficients satisfy
\begin{eqnarray}
C_{\bf 8}^Q({}^1S_0) = C_{\bf 8}^Q({}^3P_0) \equiv C_{\bf 8}^Q = 1 \,.
\end{eqnarray}
Note that there is no color-singlet operator at this order in
$\lambda$, and therefore leading Sudakov logarithms are absent in the
CSM~\cite{Hautmann:2001yz}.

\begin{figure}[t]
\centerline{\includegraphics[width=4in]{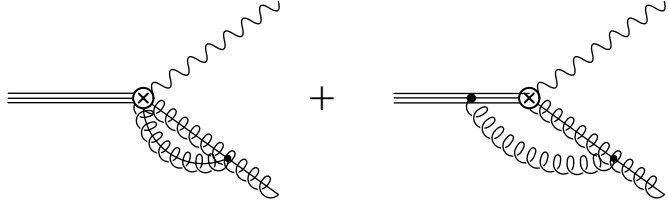}}
\caption{\it Collinear and soft diagrams needed to calculate the
renormalization of the vertices in the effective theory.}
\label{renorm}
\end{figure}

To calculate the renormalization group equations of these operators,
we also need the Feynman rules shown in Table~\ref{Feynrules}. They
are all obtained by expanding full theory diagrams in powers of
$\lambda$ and $v$. In addition the
coupling of soft gluons is given by HQET and LEET~\cite{Dugan:1991de}
Feynman rules, and the coupling of three collinear gluons is identical to
the three gluon vertex in QCD. 
\begin{table}[t]
\caption{Feynman rules in the collinear-soft theory at leading order
in $\lambda$. The vertices $V_{{\bf 1},{\bf8}}(^3S_1)$, $V_{{\bf
1},{\bf8}}(^1P_1)$ have a zero matching coefficient at this order. 
\label{Feynrules}}
\[
\begin{array}{ll}
\hbox{Diagram} & \hbox{Feynman Rule} \\
\hline \hline &\\
\epsfxsize=5cm \lower55pt\hbox{\epsfbox{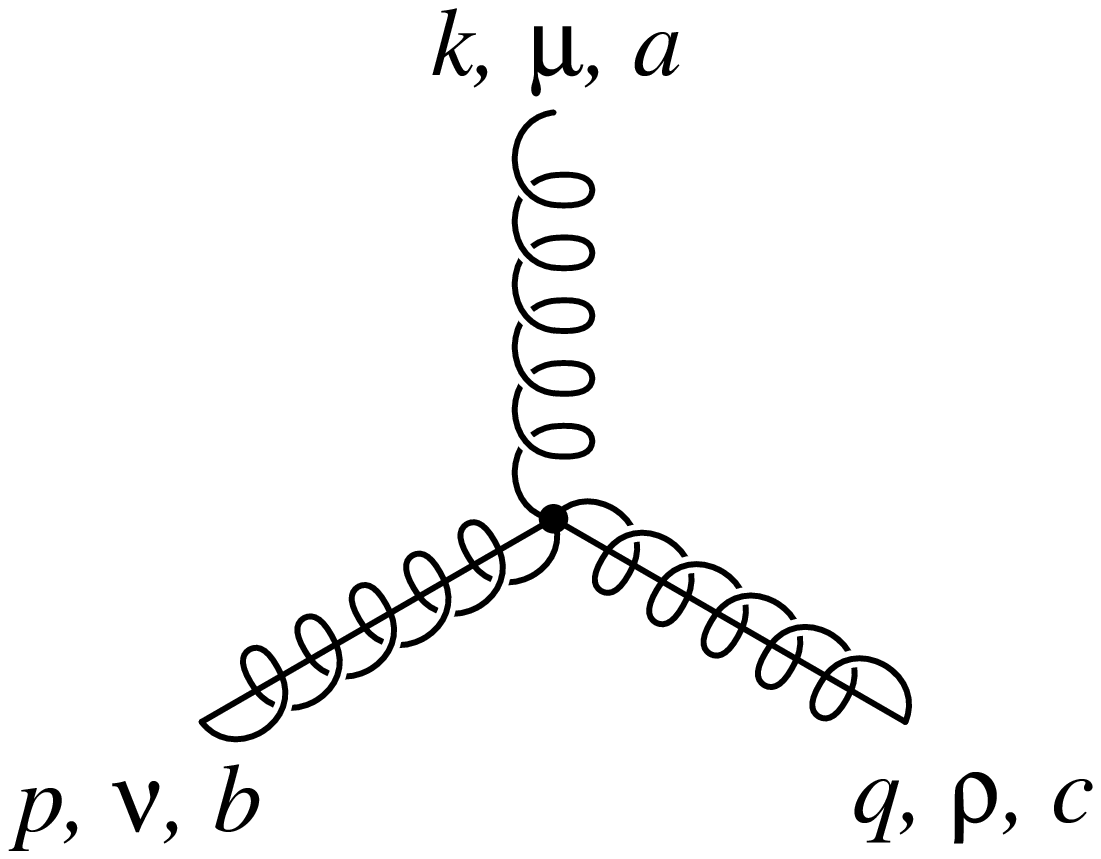}} & 
\frac{1}{2} g_s f^{abc} n^\mu (2 g^{\nu \rho} \bar{n}\cdot p -
\bar{n}^\nu p^\rho - \bar{n}^\rho p^\nu - \bar{n}^\nu \bar{n}^\rho
\bar{n}\cdot k) \\
&\\
\epsfxsize=4cm \lower35pt\hbox{\epsfbox{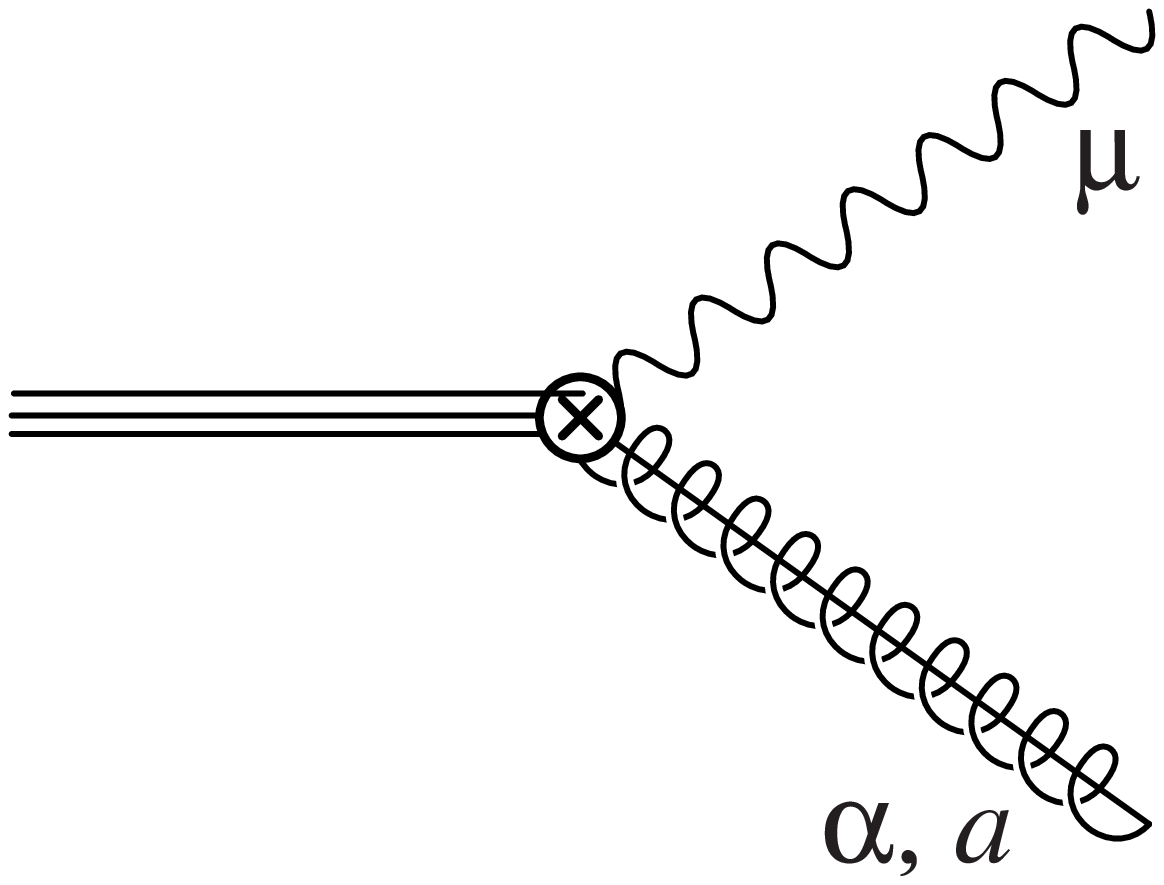}}  &
 \begin{array}{l}
   V^{\alpha\mu a}_{\bf 8}({}^1S_0) = -2 g_s e_b 
     \epsilon^{\alpha \mu}_\perp  \; \eta^\dagger_{-{\bf p}} T^a \xi_{\bf p}
   \\
   \\
   V^{\alpha\mu a}_{\bf 8}({}^3P_0) = 4 i g_s e_b (g^{\alpha
     \delta}_\perp g^{\mu\sigma} + g^{\alpha\sigma}_\perp g^{\mu\delta}
     - g^{\alpha\mu}_\perp \bar{n}^\mu \bar{n}^\delta) \; 
   \\
   \qquad\qquad\qquad\qquad\qquad
     \times\Lambda \cdot \widehat{{\bf p}}_\sigma  \eta^\dagger_{-{\bf p}}
     \Lambda\cdot\bsigma_\delta  T^a \xi_{\bf p}
  \end{array}
\\
&\\
\epsfxsize=5.0cm \lower30pt\hbox{\epsfbox{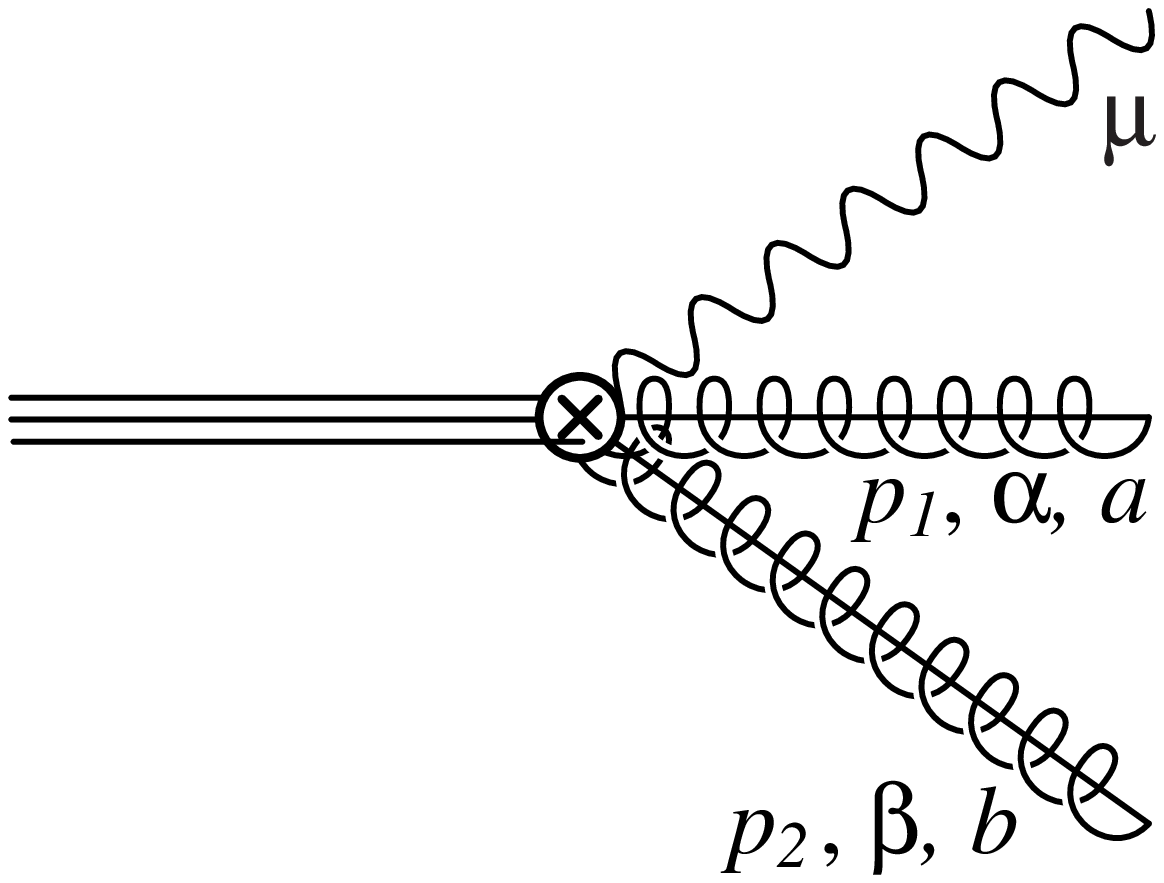}}  & 
 \begin{array}{l}
   \tilde{V}^{\alpha\beta ab}_{\bf 8}(^1S_0) = \frac{i}{2}g_s^2 e_b f^{abc}\, 
   \left(\epsilon^{\beta\mu}_\perp\frac{\bar n^\alpha}{\bar{n} \cdot p_1}- 
   \epsilon^{\alpha\mu}_\perp\frac{\bar n^\beta}{\bar{n} \cdot p_2}\right)
   \; \eta^\dagger_{-{\bf p}} T^c \xi_{\bf p}
   \\
   \\
   \tilde{V}^{\alpha\beta ab}_{\bf 8}(^3P_J) = -g_s^2 e_b\ f^{abc}\ 
   \left[\frac{\bar n^\beta}{\bar{n} \cdot p_2}
   \left(g_\perp^{\alpha\delta} g^{\mu\sigma} +
   g_\perp^{\alpha\sigma} g^{\mu\delta} -
   g_\perp^{\alpha\mu} n^\sigma n^\delta\right)\right.
   \\
   \qquad\qquad\qquad\qquad \left. 
   \,-\, \frac{\bar n^\alpha}{\bar{n}\cdot p_1}
   \left( g_\perp^{\beta\delta} g^{\mu\sigma} +
   g_\perp^{\beta\sigma} g^{\mu\delta} -
   g_\perp^{\beta\mu} n^\sigma n^\delta\right)\right]
   \\
   \qquad\qquad\qquad\qquad\qquad\qquad\times
   \Lambda\cdot\widehat{\bf p}_\sigma \; \eta^\dagger_{-{\bf p}} 
   \Lambda\cdot\bsigma_\delta T^c \xi_{\bf p}
 \end{array} 
\end{array}
\]
\end{table}
%
%\begin{figure}[t]
%\centerline{\includegraphics[width=4in]{match_sub}}
%\caption{\it Matching onto the effective field theory with a photon and 
%   two gluons.}
%\label{match_sub}
%\end{figure}
%
The renormalization is independent of the operator and the two diagrams
that need to be calculated are shown in Fig.~\ref{renorm}.
The result for the collinear and soft graphs are
\begin{eqnarray}
\label{collvertex}
{\cal A}_c &=&  \frac{ \alpha_s C_A}{4 \pi} 
\left(4 \pi \frac{\mu^2}{(-1-i \delta)p_g^2} \right)^\epsilon 
\frac{\Gamma^2(1-\epsilon) \Gamma(1+\epsilon)}{\Gamma(2-2\epsilon)}
 \frac{2-3\epsilon}{\epsilon^2} \, V_i^{\alpha\mu a} \,,\\
{\cal A}_s &=& - \frac{ \alpha_s C_A}{4 \pi} 
\left(4 \pi \frac{\mu^2 (\bar{n}\cdot p_g)^2}
{(-1-i \delta)^2 p_g^4} \right)^\epsilon 
\Gamma(1+\epsilon) \Gamma(1+2\epsilon) \Gamma(1-2\epsilon)
 \frac{1}{\epsilon^2} \, V_i^{\alpha\mu a} \,,
\end{eqnarray}
where $V_i^{\alpha\mu a} = (V^{\alpha\mu a}_{\bf 8}(^1S_0),
V^{\alpha\mu a}_{\bf 8}(^3P_J))$. 

To renormalize the vertex, we expand in $\epsilon$, keeping only the
divergent pieces.  This must equal $(Z-1) V^{\alpha\mu a}_i$. Note
however that $Z$ is not the counterterm for the vertex, rather
$Z_{\cal O}=Z_{h \bar h} Z^{1/2}_3 Z^{-1} Z_\gamma^{1/2}$, where
$Z_\gamma=1$ since we are not considering QED corrections, $Z_{h \bar
h}$ is the counterterm of the color-octet $h_v \bar h_v$ current, and
$Z_3$ is the gluon wave function counterterm:
\begin{eqnarray}
Z_{h\bar h} &=& 1+ \frac{\alpha_s C_A}{4 \pi} \frac{1}{\epsilon} \,,
\nonumber \\
Z_3 &=& 1 + \frac{\alpha_s}{4 \pi} \frac{1}{\epsilon} 
\left( C_A \frac{5}{3} -n_f \frac{2}{3} \right) \,.
\end{eqnarray}
This leads to
\begin{equation}
\label{zo}
Z_{\cal O} - 1 = \frac{\alpha_s}{4 \pi}\left[ C_A \left( \frac{1}{\epsilon^2}
+\frac{1}{\epsilon}\log\left(\frac{\mu^2}{M^2} \right)
+\frac{17}{6 \epsilon} \right) - \frac{n_f}{3 \epsilon} 
\right] \,. 
\end{equation}
We can check this result by matching the effective theory to QCD while
regulating all divergences using dimensional regularization. In this
approach there is no scale in effective theory loop integrals so they
are zero. This leaves only the counterterm $Z$, which must match the
$\epsilon$ poles in the QCD calculation. The QCD vertex at one loop
can be extracted from a calculation by Maltoni, Mangano, and Petrelli
\cite{mmp}. Equating the pole terms in the QCD calculation to $Z$ we
again obtain (\ref{zo}).

The RGE for the Wilson coefficients of the operators in
(\ref{eff_vertx}) is
\begin{eqnarray}
\mu \frac{d}{d\mu} C_i^Q(\mu) = \gamma^Q(\mu) C_i^Q(\mu)\,.
\end{eqnarray}
In order to make use of previous results from $B \to X_s \gamma$, we
write the anomalous dimension in the same form as in
Ref.~\cite{Akhoury:1996fp}
\begin{eqnarray}
\gamma^Q(\mu) = -\Gamma_{\rm cusp}^{\rm adj} \log \frac{\mu}{m_b} + 
\frac{B + 2 \gamma}{2}\,.
\end{eqnarray}
Defining
\begin{eqnarray}
\Gamma_{\rm cusp}^{\rm adj} &=& \frac{\alpha_s(\mu)}{\pi}\Gamma^{\rm 
adj}_1 + \left(\frac{\alpha_s(\mu)}{\pi}\right)^2 \Gamma^{\rm adj}_2\,, \quad
B = \frac{\alpha_s(\mu)}{\pi} B_1\,,\quad
\gamma = 
\frac{\alpha_s(\mu)}{\pi} \gamma_1,
\end{eqnarray}
we find
\begin{eqnarray}
\Gamma^{\rm adj}_1 =  C_A ,\quad 
B_1 +2 \gamma_1 = -C_A-\frac{\beta_0}{2}\,,
\end{eqnarray}
where $\beta_0 = (11C_A -2 n_f)/3$. At this point we can only
determine the above linear combination of $B_1$ and $\gamma_1$, and we
do not know the expression for $\Gamma^{\rm adj}_2$. However, as we
will show below, $B_1$ and $\Gamma^{\rm adj}_2$ can be found in the soft
theory.  To sum the leading Sudakov logarithms in the collinear-soft
theory, we run the operator given in (\ref{eff_vertx}) from the
matching scale $\mu = M$ down to a scale $\mu_c \sim \sqrt{M
\Lambda_{\rm QCD}}$ at which the OPE is performed. We can lift the
solution from Ref.~\cite{csetII}
\begin{eqnarray} \label{LOC}
 \log\bigg[\frac{C_V(\mu)}{C_V(M)}\bigg] &=& 
-\frac{4\pi\Gamma^{\rm adj}_1}{\beta_0^2\:\alpha_s(M)}
 \:\Big[ \frac{1}{y} -1 + \log y \Big] 
-\frac{\Gamma^{\rm adj}_1\beta_1}{\beta_0^3} 
\Big[ 1 -y + y\log y -\frac12 \log^2 y \Big] \nonumber\\ 
 && - \frac{B_1+2 \gamma_1}{\beta_0} \log y
  - \frac{4 \Gamma^{\rm adj}_2}{\beta_0^2}  \Big[ y -1- \log y \Big]
\,, 
\end{eqnarray}
where $y = \alpha_s(\mu)/\alpha_s(M)$.

\subsection{The purely Soft Theory}

\begin{figure}[t]
\centerline{\includegraphics[width=4in]{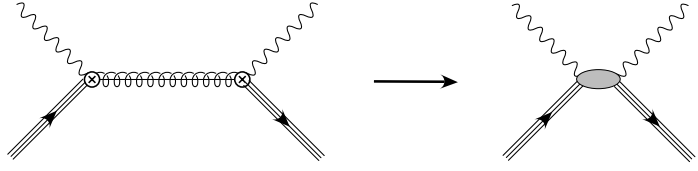}}
\caption{\it The leading OPE: tree level matching of the time ordered
product in the collinear-soft theory to a nonlocal operator in the
soft theory.
\label{leading_ope}}
\end{figure}

At the collinear scale $\mu_c \approx M\sqrt{1-z}$ we integrate out
collinear modes and perform an OPE for the inclusive $\Upsilon$
radiative decay rate in the endpoint region. The result is a nonlocal OPE in
which the two currents are separated along a light-like
direction. Diagrammatically this is illustrated in
Fig.~\ref{leading_ope}.  We write the momentum of the jet as
\begin{eqnarray}
p_X = \frac{M}{2}n^\mu + k^\mu +\frac{M}{2}(1-x)\bar{n}^\mu \,,
\end{eqnarray}
where $k^\mu$ is the residual momentum of the $b\bar{b}$ pair. Note we
distinguish $x$ from $z$, because the momentum of the jet is not
exactly the same as the momentum of the collinear gluon which was
integrated out. The two can differ slightly due to the emission of
soft quanta by the jet. The imaginary part of the tree level diagram
on the left hand side of Fig.~\ref{leading_ope} is proportional to
$\delta(p^2_x)$. Taking $k^\mu \sim M(1-x) \sim M \lambda^2$ and
expanding in the small parameter $\lambda$, we match at leading order
onto the operator
\begin{eqnarray}
\label{nonlocal_op}
{\cal O}_i(x) = 
\sum_{{\bf pp}'} [\psi^\dagger_{{\bf p}'} \Gamma_i' \chi_{-{\bf p}'}]
\,\delta(1-x+i\hat{D}^+)\,
[\chi^\dagger_{-{\bf p}} \Gamma_i \psi_{{\bf p}}] \,,
\end{eqnarray}
where $\Gamma_{\bf 8}(^1S_0)=T^a$ and $\Gamma_{\bf 8}(^3P_0) =T^a \,
{\bf p }\!\cdot\! \bsigma /\sqrt{3}$.  The $x$ serves as a continuous
label on the operator.

Each operator has a Wilson coefficient, which can depend on
the photon energy fraction $z$. The differential decay rate is 
given by a convolution of the matrix elements of the operators ${\cal O}_i(x)$
and the corresponding Wilson coefficients
\begin{eqnarray}
\label{soft_ope}
\frac{d \Gamma}{d z} = \sum_i  
\int \!d x \, C_i(x-z;\mu)f_i(x;\mu) \langle \Upsilon | {\cal O}_i | \Upsilon \rangle\,,
\end{eqnarray}
where
\begin{eqnarray}
\label{sf_def}
f_i(x,\mu) = \frac{\langle \Upsilon | {\cal O}_i(x;\mu) | \Upsilon \rangle}
{\langle \Upsilon | {\cal O}_i | \Upsilon \rangle}
\end{eqnarray}
are the lightcone distribution functions.  

The convolution of the short distance Wilson coefficients and the long
distance operators presents a technical problem since the RGE for
${\cal O}(x)$ will be given in terms of a convolution as well. We use a
Mellin transform to deconvolute (\ref{soft_ope}), which is equivalent
to taking moments of the decay rate. We must restrict ourselves to
large moments $N$, since it is the limit $N \to \infty$ which is
equivalent to the region $z \to 1$. Once the the final expression in
moment space is obtained we can take an inverse-Mellin transform to
get back to $z$-space. This procedure is valid up to corrections of
order $1-z$.  
Taking large moments of the expression in
(\ref{soft_ope}) gives
\begin{eqnarray}
\label{soft_ope_mom}
\Gamma(N) \equiv \int dz\, z^{N-1} \frac{d \Gamma}{d z} 
= \sum_i C_i(N;\mu) f_i(N;\mu)\,,
\end{eqnarray}  
where 
\begin{eqnarray}
f_i(N;\mu) \equiv \int x^{N-1} f_i(x;\mu) \,.
\end{eqnarray}

To match onto the soft theory, we compare large moments of the
differential decay rate calculated in the collinear-soft effective
theory and the soft effective theory in the parton model. Large
moments of the one loop expression calculated in the collinear-soft
theory are given in (\ref{fsme_mom}). The one loop expression for
$\langle b\bar b| {\cal O}(N;\mu) |b\bar b \rangle$ can be lifted from
Ref.~\cite{csetI} with the replacement $C_F \to C_A$:
\begin{eqnarray}
\label{xxx}
\langle b\bar b| {\cal O}(N;\mu) |b\bar b \rangle
= 1-\frac{\alpha_s C_A}{4 \pi} \left[ 4 \log^2\frac{\mu N}{Mn_0} - 4
\log\frac{\mu N}{Mn_0} \right] \,,
\end{eqnarray}
where $n_0=e^{-\gamma_E}$. At the scale $\mu_c = M \sqrt{n_0/N}$ all
logarithms match, and at that scale the tree level matching
coefficients are
\begin{eqnarray}
C_i(N;\mu) = \tilde{C}^{(0)}_i \left[C^Q(M \sqrt{n_0/N})\right]^2\,.
\end{eqnarray}

In the matrix element (\ref{xxx}) all logarithms vanish at the scale
$\mu_s=M n_0/N$.  To sum the large logarithms, we therefore have to
run the Wilson coefficient in the soft theory, $C_i(N;\mu)$, from $\mu
_c = M\sqrt{n_0/N}$ to $\mu_s = M n_0/N$. Again, we keep the notation
introduced for $B \to X_s \gamma$ and write the RGE as
\begin{equation}
\label{soft_rge}
\mu \frac{d}{d\mu}C_i(N;\mu) = \gamma(N;\mu) C_i(N;\mu),
\end{equation}
with
\begin{eqnarray}
\gamma(N;\mu) = 2 \Gamma_{\rm cusp}^{\rm adj}(\mu)\log 
\frac{\mu N}{M n_0} + B\,.
\end{eqnarray}
From the results in Refs.~\cite{csetI,Korchemsky:1987ts,KM} 
and the previous section we obtain
\begin{eqnarray}
\label{param}
\Gamma^{\rm adj}_1 &=&  C_A ,\quad 
\Gamma^{\rm adj}_2 =  
   C_A \left[ C_A \left( \frac{67}{36} - \frac{\pi^2}{12} \right) 
    - \frac{5n_f}{18} \right] , \nonumber \\
B_1 &=& -C_A\,, \quad \gamma_1 = -\frac{\beta_0}{4}\,,
\end{eqnarray}
where we have taken $\Gamma^{\rm adj}_2$ from 
Refs.~\cite{Korchemsky:1987ts,KM}.

The solution to (\ref{soft_rge}) combined with the running in the
collinear-soft theory (\ref{LOC}) can be lifted directly from
Ref.~\cite{Akhoury:1996fp} by substituting into that result the
expressions in (\ref{param}). This gives the fully resummed result in
moment space
\begin{equation}
\label{fullyresummed}
\Gamma(N)= \sum_i 
\tilde{C}^{(0)}_i f(N;M  n_0/N) e^{\log(N) g_1(\chi) + g_2(\chi)},
\end{equation}
where 
\begin{eqnarray}
\label{gis}
g_1(\chi) &=& 
-\frac{2 \Gamma^{\rm adj}_1}{\beta_0\chi}\left[(1-2\chi)\log(1-2\chi)
-2(1-\chi)\log(1-\chi)\right], \nonumber \\
g_2(\chi) &=& -\frac{8 \Gamma^{\rm adj}_2}{\beta_0^2}
  \left[-\log(1-2\chi)+2\log(1-\chi)\right] \nonumber\\
 && - \frac{2\Gamma^{\rm adj}_1\beta_1}{\beta_0^3}
   \left[\log(1-2\chi)-2\log(1-\chi)
  +\frac12\log^2(1-2\chi)-\log^2(1-\chi)\right] \nonumber\\
 &&+\frac{4\gamma_1}{\beta_0} \log(1-\chi) + 
 \frac{2B_1}{\beta_0} \log(1-2\chi) -\frac{4\Gamma^{\rm adj}_1}{\beta_0}\log n_0
 \left[\log(1-2\chi)-\log(1-\chi)\right]\,,
\end{eqnarray}
$\chi=\log (N)\, \alpha_s(M)\beta_0/4\pi$, and 
$\beta_1 = (34C_A^2-10C_A n_f-6C_F n_f)/3$.

\section{Results}

Now that we have the resummed rate in moment space
(\ref{fullyresummed}), we must take the inverse-Mellin transform to
obtain the expression for the photon energy spectrum. Fortunately, the
inverse-Mellin transform of the resummed rate in $B \to X_s \gamma$
decay has been calculated in Ref.~\cite{LLR}, and we can use that
result by simply substituting in the $g_i$ from (\ref{gis}). We obtain
the following for the resummed Wilson coefficients of the octet
contribution
\begin{eqnarray}
\label{xspaceresummed}
C_i(x-z)= -\, \tilde{C}_i^{(0)}\, \frac{d}{dz} \left\{
\theta(x-z) \; \frac{\exp [ l g_1[\alpha_s \beta_0 l/(4\pi)] +
g_2[\alpha_s \beta_0 l/(4\pi)]]}{\Gamma[1-g_1[\alpha_s \beta_0
l/(4\pi)] - \alpha_s \beta_0 l/(4\pi) g_1^\prime[\alpha_s \beta_0
l/(4\pi)]]}\right\} \,,
\end{eqnarray}
where $\tilde{C}^{(0)}_i= \tilde{C}^{(0)}_{\bf 8}(^1S_0)$,
$\tilde{C}^{(0)}_{\bf 8}(^3P_0)$, and $l \approx -\log(x-z)$. Each
$C_i(x-z)$ is evaluated at the soft scale so that all leading and
next-to-leading Sudakov logarithms have been summed into it.

One way of checking (\ref{xspaceresummed}) is by expanding in powers
of $\alpha_s$ and comparing to the fixed order calculation.  Using
\begin{equation}
\left(\frac{\log^n(1-z)}{1-z}\right)_+ = \lim_{\eta\to 0}
  \left[ \theta( 1-z -\eta) \frac{\log^n(1-z)}{1-z} +
  \delta(1-z)\, \frac{\log^{n+1}(\eta)}{n+1} \right],
\end{equation}
as the definition for plus distributions it is straightforward to
verify that the order $\alpha_s$ term in the expansion of
(\ref{xspaceresummed}) reproduces the plus distributions in
(\ref{WCoct1}).

\begin{figure}[t]
\centerline{\includegraphics[width=4in]{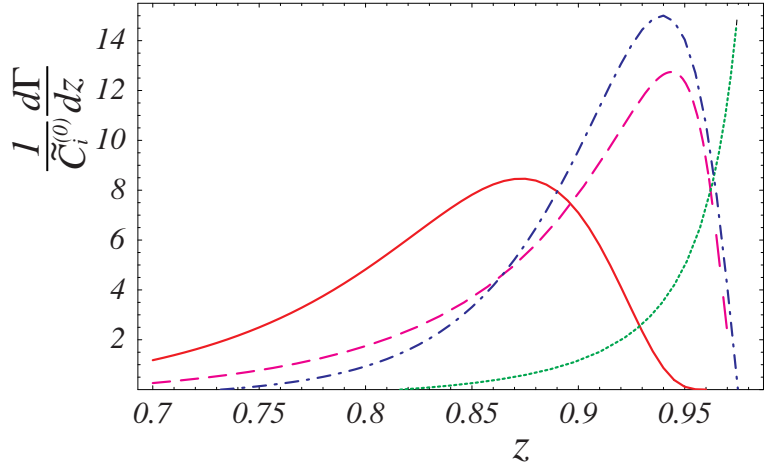}}
\vspace{.3cm}
\caption{\it The differential decay spectra near the endpoint region
$0.7<z$ in arbitrary units. The solid curve is the perturbative
resummation convoluted with the structure function and the dashed
curve is the perturbative resummation $C_i(1-z)/\tilde{C}^{(0)}_i$.
The dotted curve is the plus distribution terms in the one-loop result
(\ref{WCoct1}), and the dot-dashed curve is these terms convoluted
with the structure function.}
\label{spectrum}
\end{figure}

Recall that the covariant derivative in (\ref{nonlocal_op})
is of order $\Lambda_{\rm QCD}/M$. If we consider the limit
$1-z \gg \Lambda_{\rm QCD}/M$, then
the covariant derivative in the operator appearing in the structure function can be
neglected. In this limit we can perform the integral in (\ref{soft_ope}) to
obtain
\begin{eqnarray}
\frac{d \Gamma}{d z} = \sum_i C_i(1-z) \,.
\end{eqnarray}
 This result gives the effect of the perturbative resummation without
the structure function. The quantity $C_i(1-z)/\tilde{C}^{(0)}_i$,
which is the same for the two leading octet configurations, is shown
as the dashed line in Fig.~\ref{spectrum}.\footnote{The Landau pole
in (\ref{xspaceresummed}) should be dealt with in the same fashion as
in the $B$ decays\cite{Leibovich:2001ra}.  }

However, for the $b\bar{b}$ system $mv^2 \sim \Lambda_{\rm QCD}$ so the
covariant derivative cannot be dropped in the endpoint region of the
photon spectrum, and the differential rate is given by the
convolution in (\ref{soft_ope}). The lightcone distribution function
is a nonperturbative function and needs to be modelled. In this paper
we will be content with the simple structure functions introduced in
Ref.~\cite{Kagan:1999ym} for inclusive $B$ decays
\begin{eqnarray}
f(k^+) &=& N \left( 1-\frac{k^+}{\bar \Lambda} \right)^{a} \,
         e^{(1+a)k^+/\bar \Lambda} \;,
\label{shape1}
\end{eqnarray}
where $N$ is chosen so that the integral of the structure function is
normalized to one. In principle the structure function can be
different for the different color-octet states. But since we are
ignorant of the nonperturbative structure function, we will naively
use the same model for both the ${}^1S_0$ and ${}^3P_0$
configurations. The structure function for $B$ mesons have the
property that the first moment with respect to $k^+$ vanishes. For
quarkonium the first moment of (\ref{sf_def}) (where $x = 1+\hat k^+$)
with respect to $\hat k^+$ is
\begin{eqnarray}
\Lambda_1 = 
\frac{ \langle\Upsilon|\sum_{{\bf p}, {\bf p'}} [\chi^\dagger_{-{\bf p'}} 
T^a \Gamma_i \psi_{\bf p'}]  i \hat{D}^+ 
[\psi^\dagger_{{\bf p}} T^a \Gamma_i \chi_{-{\bf p}}]|\Upsilon\rangle }
{\langle\Upsilon| \sum_{{\bf p}, {\bf p'}} [\chi^\dagger_{-{\bf p'}} 
T^a \Gamma_i \psi_{\bf p'}][ \psi^\dagger_{\bf p} T^a 
\Gamma_i \chi_{-\bf p}]|\Upsilon\rangle }\,.
\end{eqnarray}
Therefore, we need to shift $k^+$ in (\ref{shape1}) to $k^+ +
\Lambda_1$, so that (\ref{shape1}) will have the desired first moment
for quarkonium decays.  The integration limits for $k^+$ are, similar
to the case for $B$ decays, from $-M$ to $M_\Upsilon-M$.  Both $\bar
\Lambda$ and $\Lambda_1$ are nonperturbative parameters related
through $\bar{\Lambda}=M_\Upsilon -M -\Lambda_1$.  We use the
following numbers in our plots: $\alpha_s=0.2$, $\alpha=1/137$,
$m_b=4.8 \, {\rm GeV}$, $M_{\Upsilon}=9.46 \, {\rm GeV}$, $a=1$, $\bar
\Lambda = 480\, {\rm MeV}$, and $\Lambda_1 = -620\, {\rm MeV}$.

In Fig.~\ref{spectrum} the convolution of $C_i(x-z)/\tilde{C}^{(0)}_i$
with the model of the structure function (\ref{shape1}) is shown as the
solid line. In addition we show as the dotted line the terms in the
NLO QCD expression that dominate in the endpoint region (\ref{WCoct1})
divided by $\tilde{C}^{(0)}_i$, and as the dot-dashed line the
convolution of these terms with (\ref{shape1}). Thus
Fig.~\ref{spectrum} gives a picture of the effects of resummation. The
singular plus distribution piece of the NLO QCD expression is tamed by
both the perturbative resummation and the structure function. Either
of these alone gives a similar curve, which is peaked near
$z=0.94$. However, to be consistent the perturbative resummation must
be convoluted with the structure function. This gives a curve that is
broader, with a peak that is $34\%$ lower and shifted to
$z=0.87$. Changing the values of the structure function parameters
changes the shape of the curve. Halving the value of $\bar \Lambda$
gives a narrower peak that is $30\%$ lower and shifted to $z=0.83$. If
$\alpha_s$ is increased by $10 \%$ the peak moves slightly to the left
and decreases in height by $5\%$. Doubling the value of $a$ in
(\ref{shape1}) slightly raises the peak, and steepens the curve
as it goes to zero at the endpoint.

\begin{figure}[t]
\centerline{\includegraphics[width=4in]{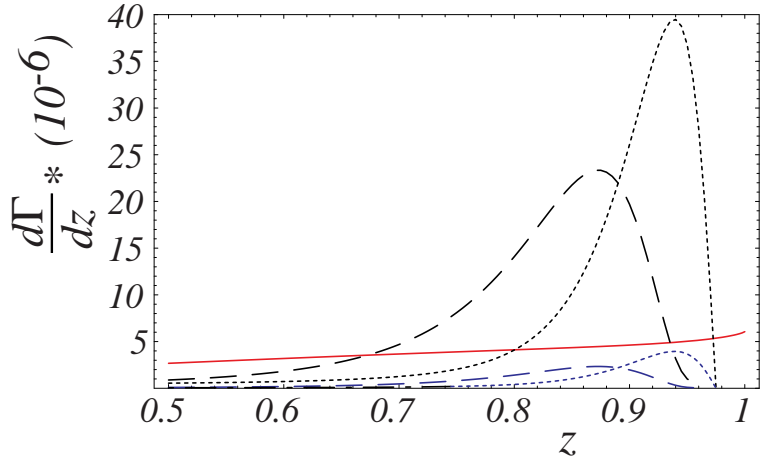}}
\vspace{.3cm}
\caption{\it The differential decay spectra in the region $0.5<z$. The
dashed curves are the fully resummed result convoluted with the shape
function for two choices of the octet matrix elements. In addition we
have interpolated the fully resummed result with the NLO result in the
region away from the endpoint. The dotted curves are the NLO result
convoluted with the structure function for two choices of the matrix
elements. The solid curve is the color-singlet contribution.}
\label{spectrumII}
\end{figure}

In Fig.~\ref{spectrumII}, we show as dashed curves the fully resummed
color-octet contribution convoluted with the structure function for
two choices of the matrix elements. We also show as the solid
curve the color-singlet contribution. Here we used 
$\langle \Upsilon | {\cal O}_{\bf 1}(^3S_1) | \Upsilon \rangle = 3.63\
{\rm GeV}^3$~\cite{Braaten:2001cm}.  The values of the color-octet
matrix elements are not well determined. One may be tempted to use a
naive power-counting argument which gives $\langle \Upsilon | {\cal
O}_{\bf 8}(^1S_0) |\Upsilon\rangle \sim \langle \Upsilon | {\cal
O}_{\bf 8}(^3P_0) |\Upsilon\rangle /m_b^2 \sim v^4 \langle \Upsilon |
{\cal O}_{\bf 1}(^3S_1) | \Upsilon \rangle$, with $v^2 \sim
0.1$. However, this gives values for the color-octet matrix elements
that are too large to be compatible with data on the ratio of hadronic
to leptonic decays~\cite{Maltoni:1999nh}. The data suggest that the
octet matrix elements are at least an order of magnitude smaller than
the naive power-counting estimate. Therefore, we use values that are
10 times and 100 times smaller than estimates from naive power
counting, with the larger matrix element yielding the higher
peak.\footnote{In Ref.~\cite{Petrelli:1998ge} it was argued that a
factor of $1/2 N_c$ should be included in a naive estimate of the
color-octet matrix elements. The larger of our two choices for this
matrix elements is of the same order of magnitude as this modified
naive estimate.}  We have added to the resummed result the NLO QCD
result with the singular terms (\ref{WCoct1}) subtracted off. This
interpolates between the NLO QCD expression at lower values of $z$ and
the resummed result in the endpoint region. For comparison we show as
the dotted curves the NLO QCD contribution~\cite{Maltoni:1999nh}
convoluted with the shape function for the two choices of the
color-octet matrix elements.

Since the octet contributions dominate the color-singlet contribution
in this endpoint region (or at least are of the same order of
magnitude) it should be possible to use this process to constrain the
size of the color octet matrix element. The suppression of the matrix
element compared to the naive power counting estimate makes a
measurement of this matrix element particularly interesting and might
shed some light on the convergence of the $v$ expansion in NRQCD.
However, before a meaningful comparison of theory to data on radiative
$\Upsilon$ decay can be made, subleading Sudakov logarithms in the
color-singlet contribution must be summed. The existence of these
logarithms was first pointed out in Ref.~\cite{Hautmann:2001yz} where
it was observed that though leading Sudakov logarithms cancel in the
color-singlet contribution to the differential rate, they are present
in the derivative of the rate at the endpoint. We leave the
resummation of these logarithms to a future publication~\cite{future}.

\section{conclusion}

Using an effective field theory approach, we have resummed Sudakov
logarithms in the leading color-octet contributions to the $\Upsilon
\to X \gamma$ differential decay rate in the endpoint region. This is
done in two steps.  First we match onto an effective theory with
collinear and soft degrees of freedom and run the theory to the
collinear scale.  Next we integrate out collinear modes by performing
an OPE, matching onto non-local operators which are run to the
soft scale. This sums all Sudakov logarithms into Wilson coefficients
of these operators. The color-octet contribution to the differential
decay rate in the endpoint region is given by the convolution of the
Wilson coefficients with matrix elements of the operators between
$\Upsilon$ states. The latter are the color-octet structure functions
defined in Ref.~\cite{upsStructfun}.

We choose a simple model for the structure functions to investigate
the phenomenological consequences of resummation. We find that either
the perturbative resummation or the inclusion of a structure function
cures the singular behavior of the QCD results. Both give a similar
effect, causing the curve to turn over near $z=0.94$ and to go to zero
at the endpoint. However, the effective field theory approach makes it
clear that the correct expression for the differential rate near the
endpoint is given by a convolution of the perturbative resummation and
the structure function. This gives a spectrum that has a broader and
lower peak than we obtain by including only perturbative resummation
or the structure function, shifting the peak to $z=0.87$.

Before a meaningful comparison to data can be made the color-singlet
result must be resummed as well. Only then can data on the decay spectrum
be used to constrain the size of the octet contribution in the
endpoint region~\cite{future}.

\acknowledgments We would like to thank Ira Rothstein, Iain Stewart,
and George Sterman for helpful discussions. This work was supported in
part by the Department of Energy under grant numbers
DOE-FG03-97ER40546, DOE-ER-40682-143 and
DE-AC02-76CH03000. A.K.L. thanks the theory group at CMU for its
hospitality, and S.F. thanks the theory group at UCSD for its
hospitality.

\appendix

\section{Forward scattering matrix element}\label{appforward}
\label{fsme_app}

Matching the forward scattering matrix element in the QCD and
collinear-soft theory gives an important check that we are reproducing
the infrared physics of the full theory. We have already calculated
the vertex corrections. Note that there are two contributions from the
vertex loops so the vertex contribution to the forward scattering
matrix element is twice that given in (\ref{collvertex}). There is a 
contribution to the forward scattering matrix element from a ladder
graph, which we have not evaluated yet. It gives 
\begin{equation}
{\cal M}_\ell =  \frac{ \alpha_s C_A}{2 \pi} 
\left(4 \pi \frac{\mu^2(\bar{n} \cdot p_g)^2}
{(-1-i \delta)^2 p_g^4} \right)^\epsilon 
\Gamma(1+\epsilon) \Gamma(1+2\epsilon) \Gamma(1-2\epsilon)
 \frac{1}{\epsilon} \, {\cal M}_0 \,,
\end{equation}
where 
\begin{eqnarray}
{\cal M}_0 = \tilde{C}^{(0)}_i \frac{M^2}{p^2_g-i \delta}
\end{eqnarray}
is the tree level amplitude, and $p^2_g=M^2 (1-z)$. 
In addition there is a contribution coming from virtual corrections to
the collinear gluon propagator, which include the fermion, gluon, 
and ghost loops,
\begin{equation}
{\cal M}_{Z_3}= \frac{\alpha_s}{4\pi}\left(\frac53 C_A - \frac23 n_f\right)
   \left(\frac{\mu^2}{(-1-i\delta) p_g^2}\right)^\epsilon \frac1{\epsilon}\,
  {\cal M}_0 \, .
\end{equation}
Adding all contributions gives the full one-loop corrections to
the forward scattering amplitude
\begin{eqnarray}
\label{fsme_loop}
{\cal M}_{\rm lg} &=& 
\frac{ \alpha_s}{2 \pi}\left\{ C_A \left[ \frac{1}{\epsilon^2}
+\frac{1}{\epsilon}\log\left(\frac{\mu^2}{M^2} \right)
+\frac{17}{6 \epsilon} \right] - \frac{n_f}{3 \epsilon} 
+ \frac{C_A}{2} \log^2\left( \frac{\mu^2}{M^2} \right) 
+ \frac{17 C_A}{6} \log \left( \frac{\mu^2}{M^2} \right)
\right.  
\nonumber \\
&& \qquad - \left. \frac{n_f}{3} \log \left( \frac{\mu^2}{M^2} \right)
- C_A \log^2 [(1-z)(-1-i \delta)] - C_A\frac{23}{6} \log [(1-z)(-1-i \delta)]
\right.  
\nonumber \\
&& \qquad + \left.
\frac{n_f}{3} \log [(1-z)(-1-i \delta)] 
\right\} {\cal M}_0 + \dots \,,
\end{eqnarray}
where the ellipses represent terms of order $\alpha_s$ not enhanced by
logarithms.

Next add the counterterms. There is a gluon propagator insertion that
gives $-{\cal M}_0 (Z_3-1)$. The vertex insertion is $Z_{h\bar{h}} Z^{1/2}_3
Z^{-1}_O$. In order to facilitate the comparison with the results in
Ref.~\cite{Maltoni:1999nh}, we use dimensional regularization to regulate the 
infrared divergences in the heavy quark sector, and therefore $Z_{h\bar{h}}=1$.
Adding all the counterterms cancels the
$\epsilon$ poles in (\ref{fsme_loop}), leaving
\begin{eqnarray}
\label{fsme}
{\cal M}_{\rm lg} &=& 
\frac{ \alpha_s}{2 \pi}\left\{ C_A \left[ 
\frac{1}{2} \log^2\left( \frac{\mu^2}{M^2} \right) 
+ \frac{17}{6} \log \left( \frac{\mu^2}{M^2} \right) \right]
- \frac{n_f}{3} \log \left( \frac{\mu^2}{M^2} \right)
- C_A \log^2 [(1-z)(-1-i \delta)]
\right. \nonumber \\ 
&& \qquad - \left. C_A\frac{23}{6} \log [(1-z)(-1-i \delta)]
+ \frac{n_f}{3} \log [(1-z)(-1-i \delta)]
\right\} {\cal M}_0  + \dots \,.
\end{eqnarray}
To obtain the differential decay rate we take 
$-1/\pi \; {\rm Im}$(\ref{fsme}) using
\begin{eqnarray}
\frac{-1}{\pi}{\rm Im} \frac{\log[(1-z)(-1-i \delta)]}{1-z+i \delta} &=& 
\left( \frac{1}{1-z} \right)_+ \,,
\nonumber \\
\frac{-1}{\pi}{\rm Im} \frac{\log^2 [(1-z)(-1-i \delta)]}{1-z+i \delta} &=& 
2 \left( \frac{\log(1-z)}{1-z} \right)_+ - \pi^2 \delta(1-z) \,.
\end{eqnarray}
Comparing to Ref.~\cite{Maltoni:1999nh} (taking the limit $z \to 1$ in
their expressions), we confirm that at the matching scale $\mu = M$
the effective theory reproduces the plus distributions in the full
theory. Taking large moments of the imaginary part of (\ref{fsme})
gives:
\begin{eqnarray}
\label{fsme_mom}
-\frac{1}{\pi} \int dz \; z^{N-1} \; {\rm Im}{\cal M}_{\rm lg} &=& 
\frac{ \alpha_s}{2 \pi}\left\{ C_A \left[ 
\frac{1}{2} \log^2\left( \frac{\mu^2}{M^2} \right) 
+ \frac{17}{6} \log \left( \frac{\mu^2}{M^2} \right) \right]
- \frac{n_f}{3} \log \left( \frac{\mu^2}{M^2} \right)
\right. \\ 
&& \qquad \left. 
- C_A \log^2 \left[ \frac{N}{n_0} \right]
+ C_A\frac{23}{6} \log \left[ \frac{N}{n_0} \right]
- \frac{n_f}{3} \log \left[ \frac{N}{n_0} \right]
\right\} \tilde{C}^{(0)}_i + \dots \,.
\nonumber
\end{eqnarray}

%\begin{figure}[!t]
 %\centerline{\mbox{\epsfysize=4.0truecm
 % \hbox{\epsfbox{ladder.ps}} \hspace{1.cm}}} \medskip {\tighten
 % \caption[1]{The ladder graph for the forward scattering amplitude.}  
%\label{laddergraph} }
%\end{figure}

%%%%%%%%%%%%%%%%%%%%%%%%%%%%%%%%%%%%%%%%%%a
%Bibliography

\end{document}